\documentclass[a4paper,11pt]{article}
\usepackage{pos}
\newcommand{\ttbar}{t\bar{t}}

\title{Optimizing b-Jet Performance in the CMS High-Level Trigger with Run-3 Data}
\ShortTitle{b-Jet Performance in the CMS HLT with Run-3 Data}

\author*[a]{Uttiya Sarkar}
\author[]{on behalf of the CMS Collaboration}

\affiliation[a]{III. Physikalisches Institut A, RWTH Aachen,\\
  Sommerfeldstr. 16, 52074 Aachen, Germany}

\emailAdd{uttiya.sarkar@cern.ch}

\abstract{The real-time identification and selection of b-jets play a crucial role in the CMS experiment, particularly in searches involving heavy-flavor jets. The High-Level Trigger (HLT) is designed to efficiently select events of interest while maintaining a manageable output rate of a few kilohertz. This report presents the commissioning and performance evaluation of b-jet triggers in the CMS HLT system using proton-proton collision data collected during Run 3 (2022–-2024). Key aspects include algorithm optimization, efficiency studies, and comparisons with offline reconstruction. The results provide valuable insights into the current b-jet selection strategy and highlight potential refinements for future data-taking campaigns.}

\FullConference{13th Edition of the Large Hadron Collider Physics Conference (LHCP 2025)\\
04-09 May 2025\\
Taipei, Taiwan\\}

\begin{document}
\maketitle

\section{Introduction}

The identification of jets originating from heavy quarks, namely b- and c-jets, is a cornerstone of the CMS physics program~\cite{CMS:2017wtu}. Efficient heavy-flavor tagging at the trigger level is particularly important in fully hadronic final states, where no other distinctive features (such as high-$\mathrm{p_T}$ leptons or large missing transverse momentum) are available. This is crucial for measurements of Standard Model (SM) processes, e.g.\ $t\bar{t}H$~\cite{CMS:2018hnq,Marchegiani:2024yhi} and vector-boson-fusion~\cite{CMS:2023tfj}, as well as for searches like Higgs boson pair production ($H\rightarrow b\bar{b}$)~\cite{CMS:2018nsn,CMS:2020zge,PhysRevLett.129.081802,PhysRevLett.122.121803} and new resonances\cite{CMS:2024phk,CMS:2016fse}.  

\begin{figure}[!htb]
  \centering
  \includegraphics[width=0.6\textwidth]{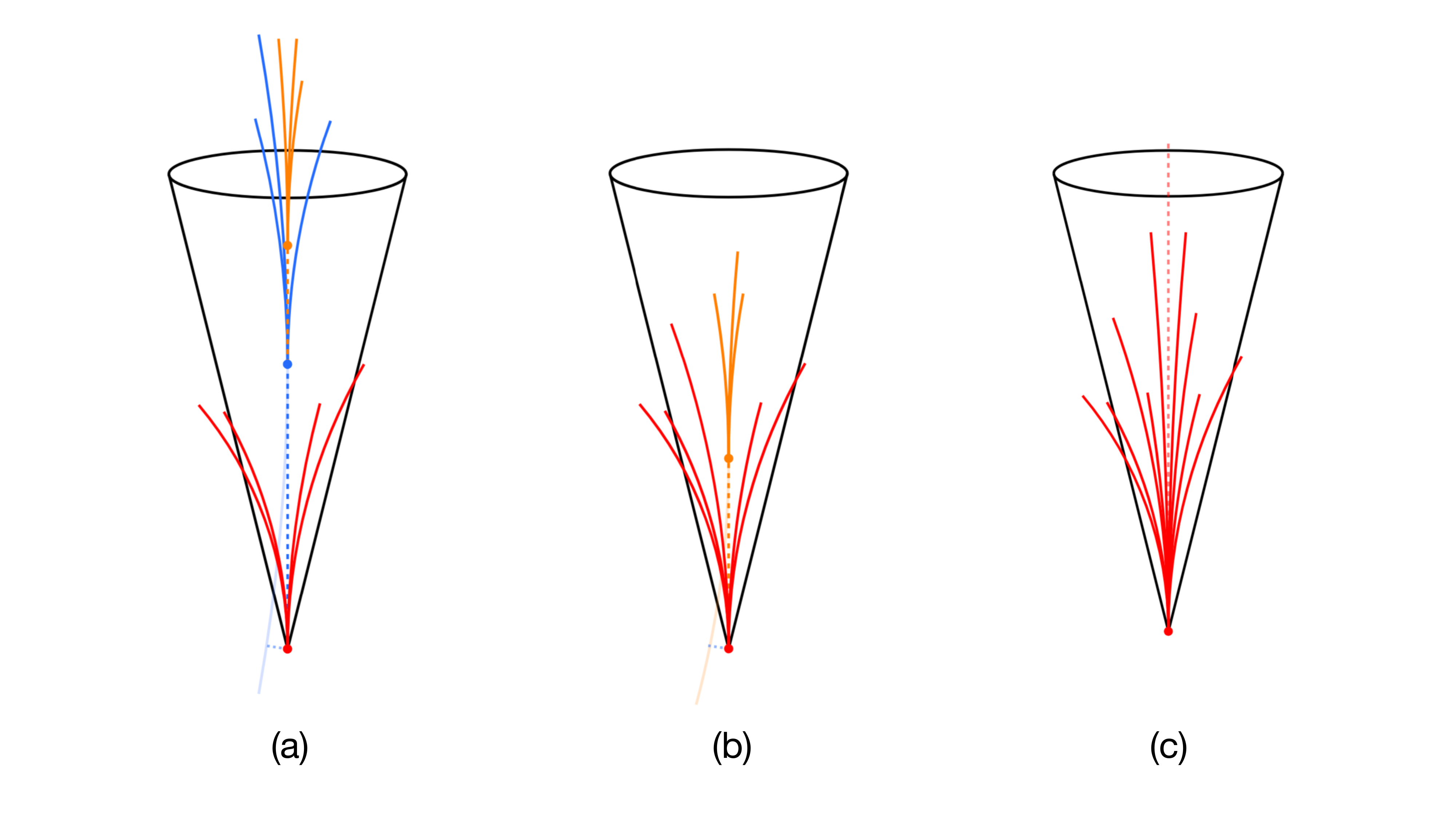}
  \caption{Illustration of jets initiated by (a) b-quark, (b) c-quark, and (c) light quark or gluon. Secondary vertices (SV) from heavy-flavor hadron decays are displaced from the primary vertex (PV), providing a key handle for heavy-flavor jet tagging at the trigger level.}
  \label{fig:01}
\end{figure}

Discriminating b- jets from light-flavor (udsg) or c-jets exploits both the jet fragmentation pattern and the relatively long lifetime of B-hadrons. Their decays typically produce a displaced secondary vertex (SV) and tracks with large impact parameters with respect to the primary vertex (PV). Semileptonic decays, present in about 20\% of b-hadron decays, provide an additional discriminating handle. Figure~\ref{fig:01} illustrates schematic differences between b, c, and light-flavor jets.  

During Run--2, the heavy-flavor triggers based on multivariate classifiers (fully connected CNN+RNN), namely \textsc{DeepJet}~\cite{Bols_2020} were deployed. However, under the aged tracker conditions of Run--3, these approaches could not be operated at sustainable rates without raising thresholds. To overcome this, ParticleNet~\cite{Qu:2019gqs} is introduced. It is a graph neural network optimized and trained for jet flavor tagging dedicated for the High-Level Trigger (HLT). This algorithm provides significant gains in b-jet efficiency while reducing mistag rates, and for the first time allows dedicated c and $\tau$ jet triggers from 2023 onwards. These improvements are central to unlocking hadronic final states in Run--3 analyses.  

In this report, we briefly discuss the CMS HLT design and the Run--3 update from the \textsc{DeepJet} tagger to ParticleNet@HLT. We then present the ParticleNet@HLT performance with 2024 data and b-tagging results in double b-jet and the dedicated $HH \to 4b$ physics trigger paths.

\section{The CMS High-Level Trigger}
The CMS trigger system reduces the LHC collision rate of $\sim$30~MHz to about 1--3~kHz suitable for offline storage. This is achieved in two stages: the hardware-based Level-1 (L1) trigger with a latency of $\sim$4~$\mu$s and an output rate of $\mathcal{O}(100$~kHz), followed by the software-based High-Level Trigger (HLT) running on a large processor farm~\cite{CMS:2024aqx}.  

The HLT is organized into paths that reconstruct physics objects and apply kinematic and quality requirements tailored to specific signatures. Early selections use fast calorimeter and muon information, while more complex algorithms such as particle-flow (PF) reconstruction~\cite{CMS:2017yfk}, tracking, and heavy-flavor tagging are deferred to later stages to optimize resources. During Run--3, an upgraded  track reconstruction algorithm\cite{Bruschini:2025qba} enabled a higher HLT efficiency and reduced latency compared to Run--2. Physics objects are then reconstructed with the PF algorithm and clustered into jets with the anti-$k_T$ algorithm. These jets form the basis for b- and c-tagging at the trigger level.  

\section{ParticleNet@HLT for b-Jet Triggers}
During early Run--2, online b-tagging of anti-$k_T$ jets with a cone radius of 0.4 (AK4 jets) was performed with the \textsc{DeepCSV} algorithm~\cite{CMS-DP-2017-005}, a feed-forward DNN trained on track- and vertex-related features. During early Run--3, \textsc{DeepJet}~\cite{CMS-DP-2022-030} and ParticleNet~\cite{CMS-DP-2023-021} were introduced as new classifiers for heavy-flavor tagging at the HLT. Early in 2022, several b-jet triggers employed \textsc{DeepJet}, but after validation it was found that ParticleNet significantly outperformed both \textsc{DeepJet} and the Run--2 \textsc{DeepCSV} tagger. From 2023 onwards, ParticleNet became the standard online b-tagger in the CMS HLT.  

ParticleNet is a dynamic graph convolution neural network~\cite{10.1145/3326362} that represents a jet as an unordered set of its PF candidates and secondary vertices. Relations among these constituents are captured through successive edge-convolution layers, enabling the network to learn both local and global correlations in a permutation-invariant way. The HLT-optimized ParticleNet@HLT model uses track impact parameters, flight-distance observables, SV kinematics, and jet composition variables as inputs, with displaced tracks and SV features providing the most powerful discriminants for b-jet identification.  

Training is performed on simulated AK4 jets with $p_T>30$~GeV and $|\eta|<2.5$, matched to generator-level jets. The model predicts per-jet probabilities for five classes: b, c, light (uds), gluon, and hadronic $\tau$. Dedicated working points are defined to target c and light (udsg) misidentification rates of 10\%, 1\%, and 0.1\% (loose, medium, and tight).  

\begin{figure}[!htb]
    \centering
    \includegraphics[width=0.5\textwidth]{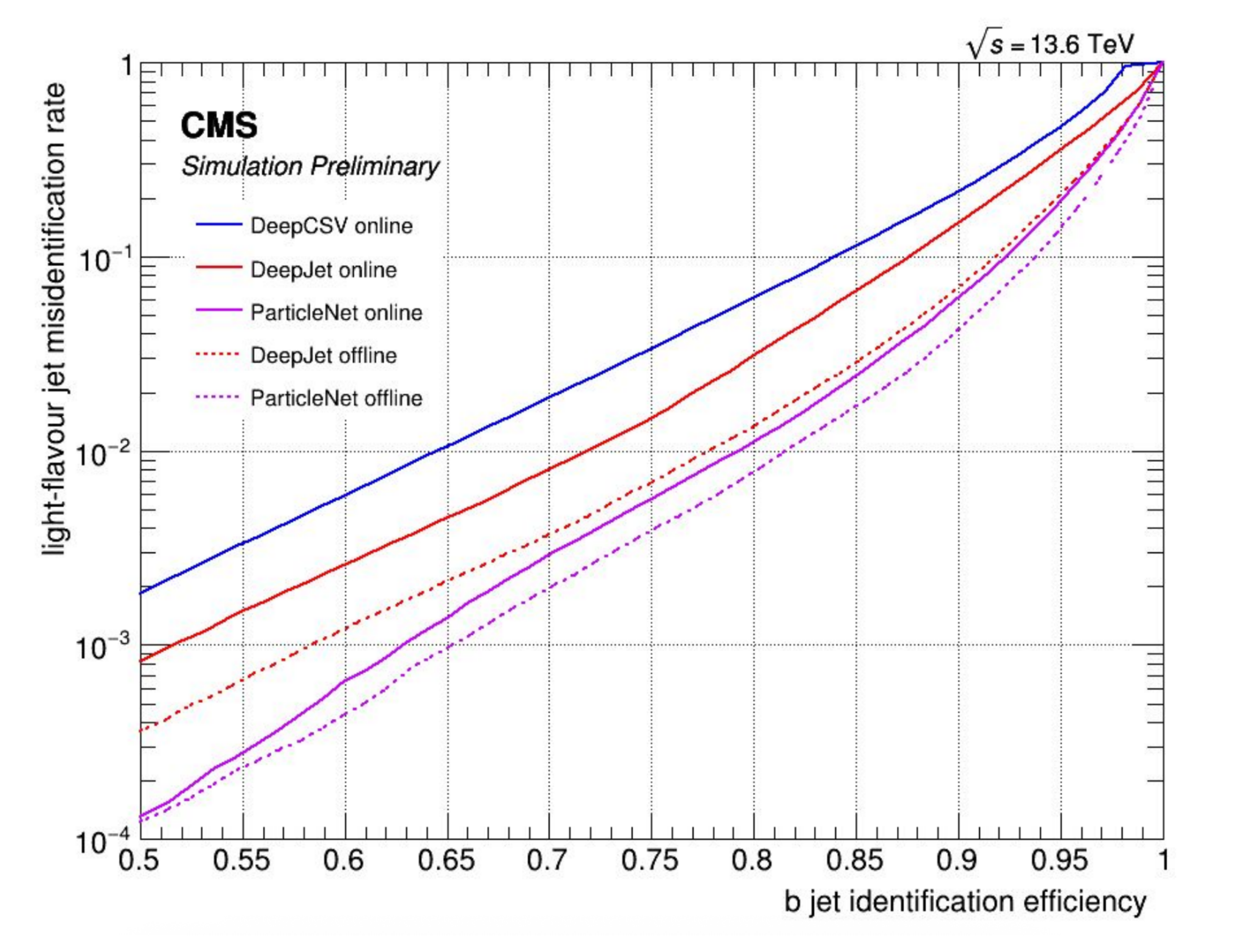}
    \caption{
    Performance of the ParticleNet@HLT algorithm for AK4 b-jet tagging at the HLT.
    Shown is the b-jet tagging efficiency versus the misidentification probability
    for c and light-flavor jets in simulated $\ttbar$ events with $\mathrm{p_{T}} > 30$~GeV
    and $|\eta| < 2.5$. Results are compared to the Run--3 \textsc{DeepJet} and Run--2 \textsc{DeepCSV}
    taggers. ParticleNet demonstrates a $\sim$10--15\% gain in efficiency at fixed mistag rates,
    representing the state-of-the-art for online AK4 b-tagging in CMS.
    }
    \label{fig:02}
\end{figure}

As shown in Fig.~\ref{fig:02}, ParticleNet@HLT achieves a $\sim$10–15\% gain in b-tagging efficiency at fixed misidentification rates compared to \textsc{DeepJet}, with even larger improvements over \textsc{DeepCSV}. 

\section{b-tagging performance on data}
\subsection{ParticleNet@HLT performance in 2024 data}
ParticleNet@HLT model deployed in the HLT menu processes the HLT jets in real time and assigns b-tagging scores. Three working points (loose, medium, tight) are defined by thresholds on the ParticleNet score, corresponding to misidentification probabilities of 10\%, 1\%, and 0.1\% for c- and light-flavor jets. 

\begin{figure}[!htb]
    \centering
    \includegraphics[width=0.32\textwidth]{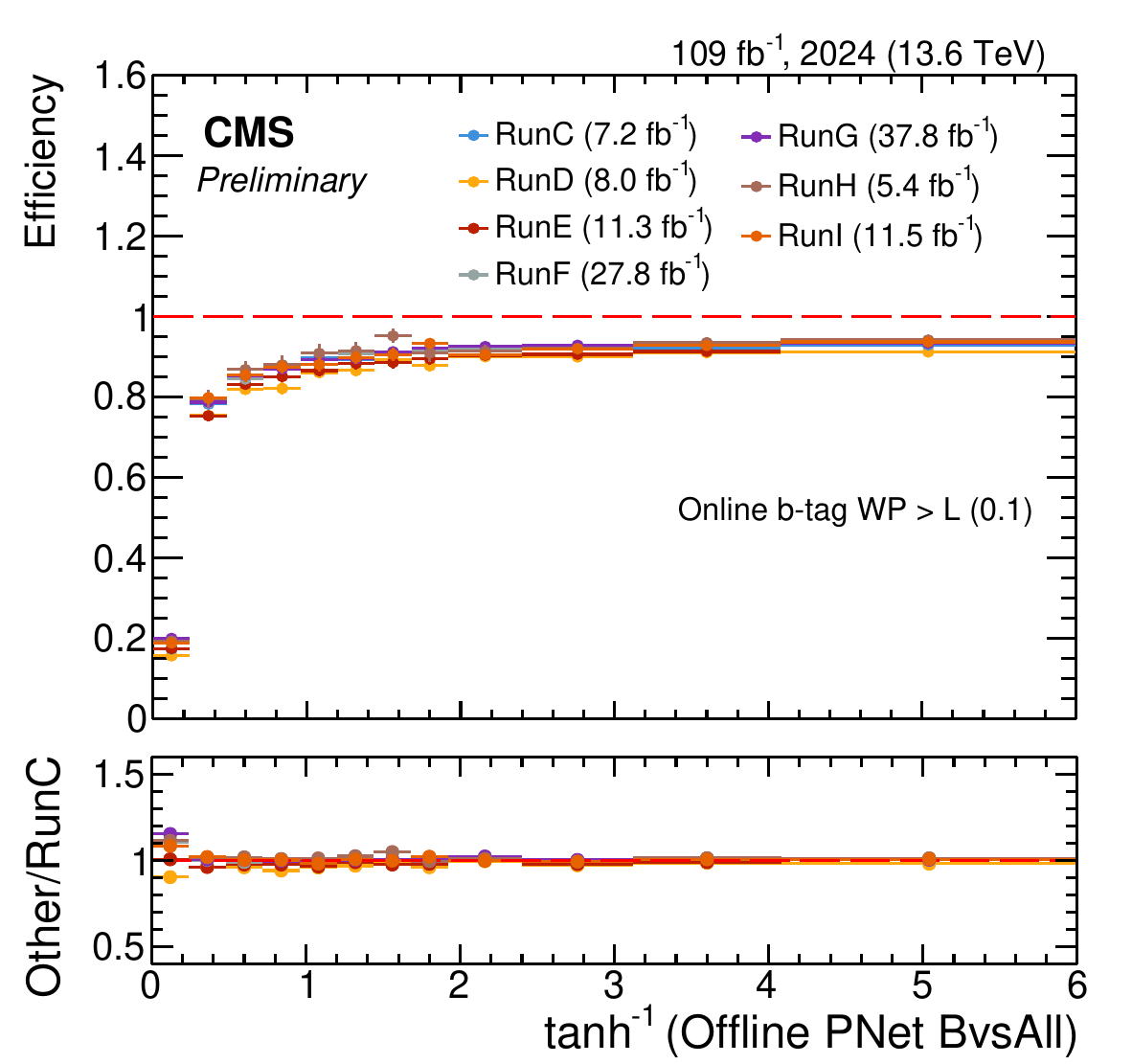}
    \includegraphics[width=0.32\textwidth]{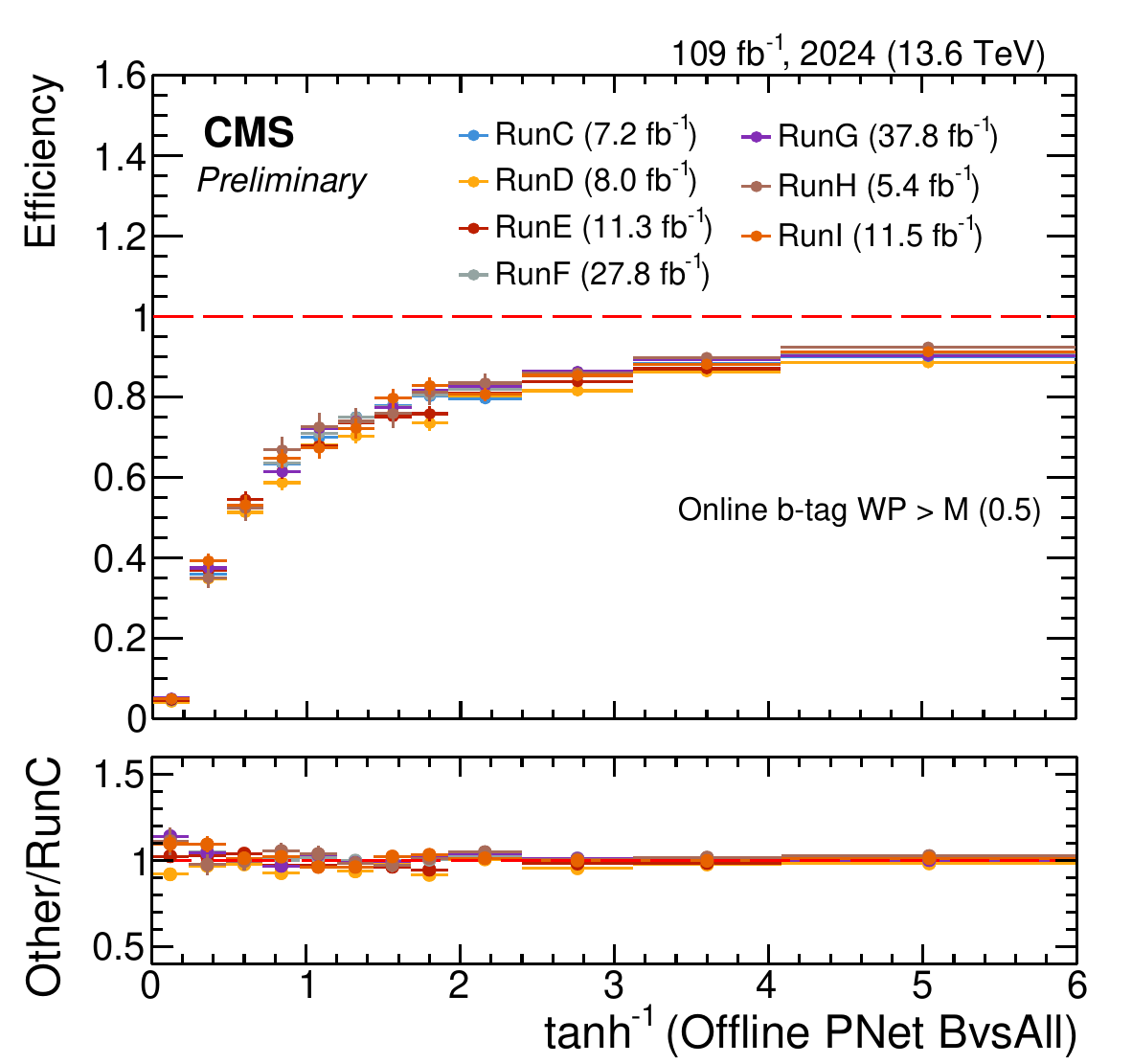}
    \includegraphics[width=0.32\textwidth]{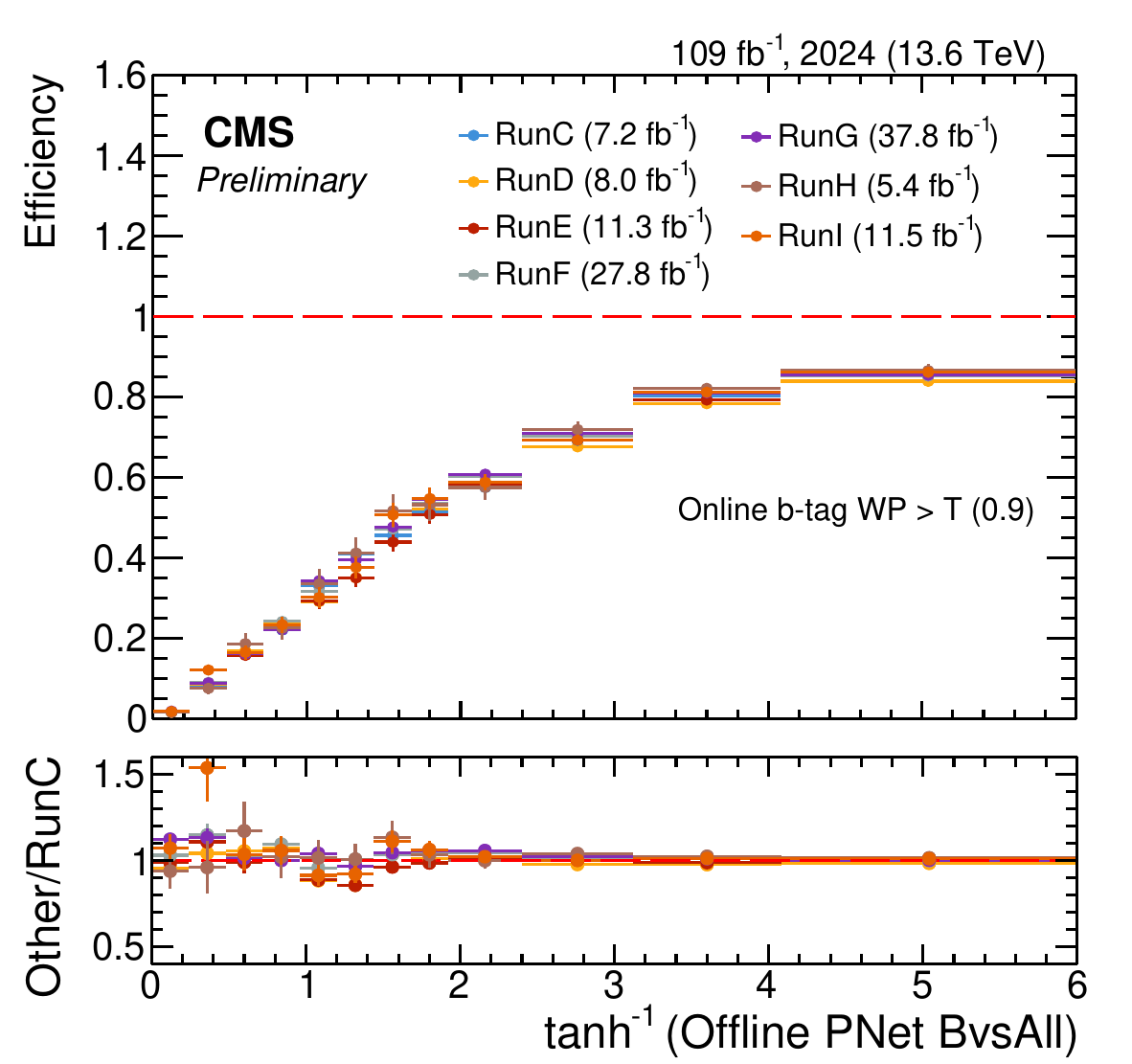}
    \caption{Per-jet ParticleNet@HLT efficiency vs.\ transformed offline ParticleNet BvsAll score for 2024 data. Plots from left to right show the loose (L), medium (M), and tight (T) online WPs. The top panel compares the efficiency across different data-taking eras, from RunC to RunI.I n the bottom panel, RunC is used as the reference, and the ratios of all other run eras are shown normalized to it.  
}
    \label{fig:03}
\end{figure}

The algorithm was deployed online in 2022 and retrained in 2023 using updated simulated samples enriched in heavy-flavor processes. Performance studies are carried out in a $t\bar{t}$-enriched dataset, requiring at least two offline jets with $\mathrm{p_T} > 35$~GeV and $|\eta| < 2.5$, matched to online jets with $\mathrm{p_T} > 30$~GeV and $|\eta| < 2.5$~\cite{CMS-DP-2025-013}. The per-jet efficiency is defined as
\[
\epsilon_{\text{jet}} \;=\;
\frac{N_{\text{jets}}\big(\text{offline jets matched to online jets passing ParticleNet@HLT WP}\big)}
{N_{\text{jets}}\big(\text{offline jets matched to online jets}\big)}.
\]
Figure~\ref{fig:03} shows the measured per-jet efficiency in 2024 data as a function of the transformed offline ParticleNet@HLT discriminator score (\texttt{BvsAll} = {\texttt{prob(b)}} / {[1-\texttt{prob(b)}]}), divided into different data-taking eras (from RunC to RunI). The plot demonstrates stable performance throughout the data-taking period. The three plots correspond to the loose, medium, and tight online WPs respectively; the lower subpanels show era-by-era ratios relative to RunC.

ParticleNet@HLT demonstrated a stable operation throughout 2024.

\subsection{Online b-tagging performance in physics triggers}
Trigger-level performance is evaluated using physics selections and therefore uses a different efficiency definition. The ParticleNet@HLT trigger efficiency is measured in a $\ttbar$+jets (electron–muon) control region and is defined as
\[
\epsilon_{\text{trig}} \;=\;
\frac{N_{\text{events}}\big(\text{HLT } e\mu + 2b \text{ path with ParticleNet@HLT}\big)}
{N_{\text{events}}\big(\text{offline } e\mu + 2b \text{ selection}\big)}.
\]
Here the offline denominator requires an opposite-sign, well-identified and isolated $e\mu$ pair and two offline b-tagged jets with $p_T>30$~GeV and $|\eta|<2.5$ (offline ParticleNet WP at $\sim$84\% efficiency, $\sim$1\% mistag)~\cite{CMS-DP-2025-009}.

\begin{figure}[!htb]
    \centering
    \includegraphics[width=0.45\textwidth]{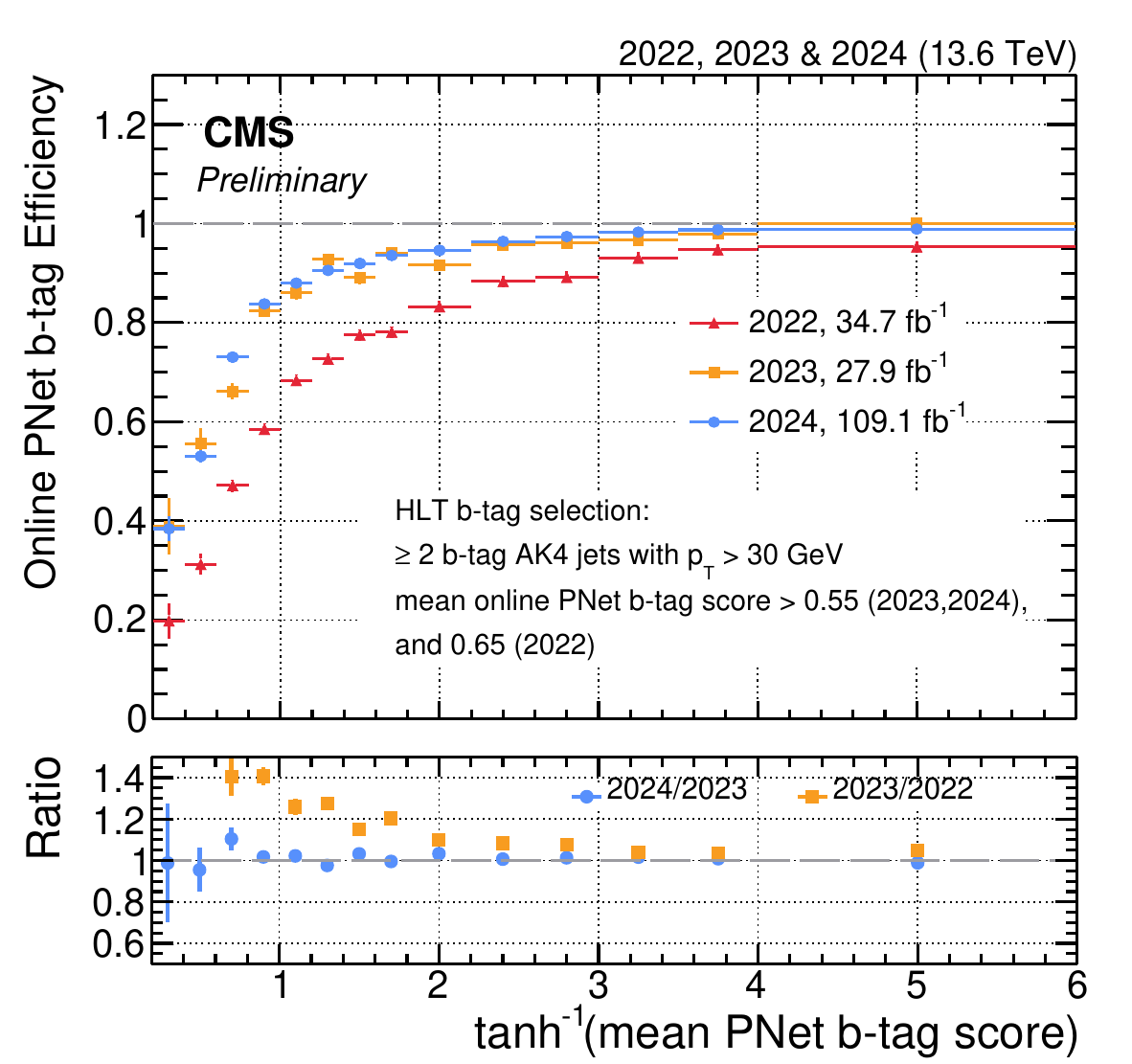}
    \includegraphics[width=0.45\textwidth]{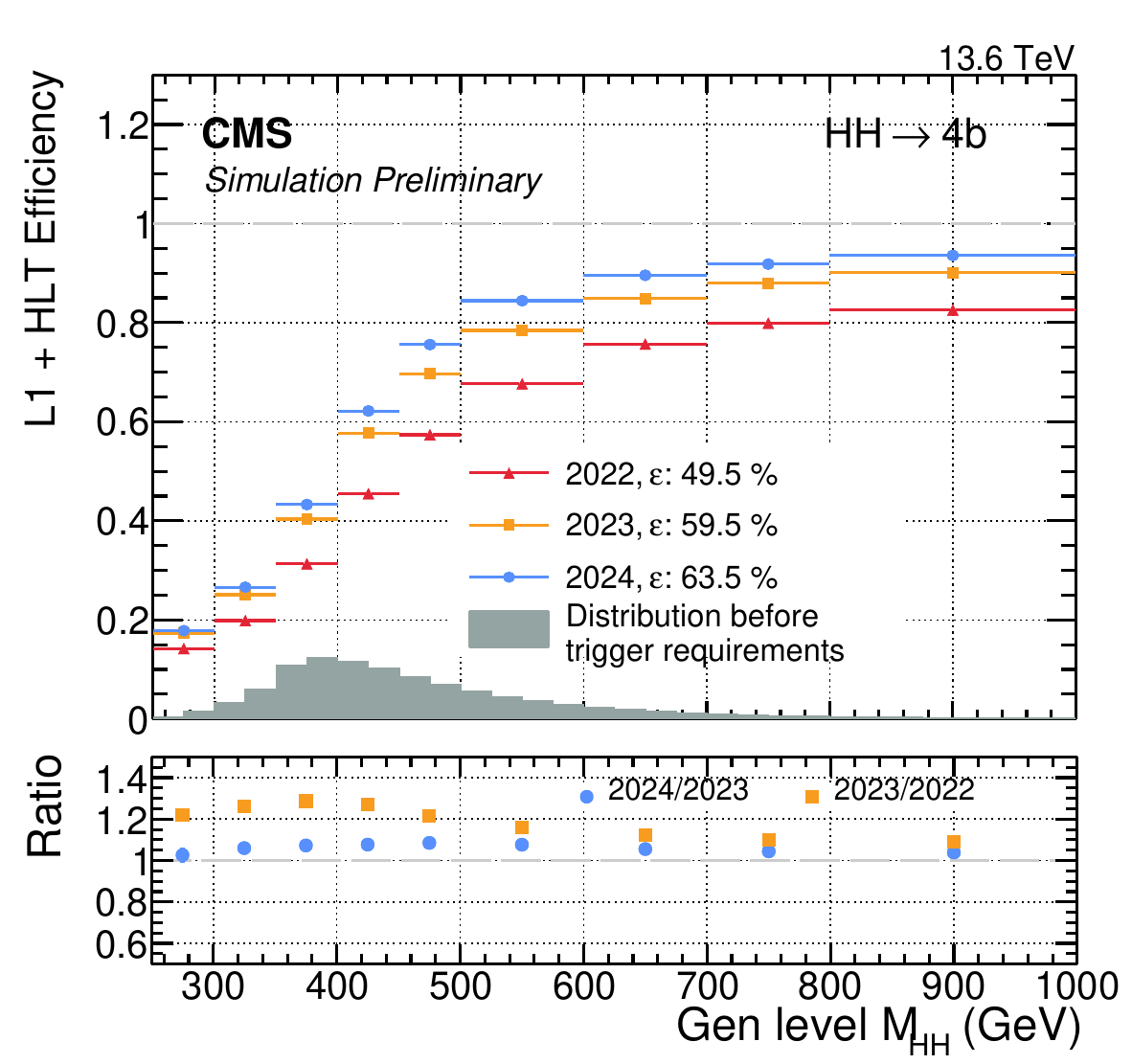}
    \caption{Left: Trigger efficiency vs.\ mean offline ParticleNet score in the $e\mu$ control region for 2022 (red), 2023 (orange), and 2024 (blue). 
    Right: $4j2b$ trigger efficiency on simulated $HH\to 4b$ events vs.\ $M_{HH}$ for 2022–2024.}
    \label{fig:04}
\end{figure}

Figure~\ref{fig:04} (left) presents the trigger efficiency as a function of the mean offline ParticleNet score of the two leading b-tagged jets for 2022–2024 collision data (year-to-year comparison). Figure~\ref{fig:04} (right) shows the L1+HLT efficiency of the $4j2b$ trigger, which requires at least four jets and two of them b-tagged at the HLT, on simulated $HH\to 4b$ events as a function of generator-level $M_{HH}$ for the same years; lower panels display relative gains due to the retraining and adjustments of WPs in 2023 (2023 vs 2022 and 2024 vs 2023). The 2024 HLT menu (with looser online selection) yields an approximate $\sim$6\% efficiency improvement over 2023 for the $4j2b$ topology, with an associated rate increase of about $50\,$Hz.

\section{Conclusions}
ParticleNet@HLT has markedly improved online b-jet identification in the CMS HLT during Run--3, 
delivering higher efficiency and stable performance across all data-taking periods. Regular 
retraining with updated datasets has been key to maintaining this robustness. Looking ahead, 
the adoption of transformer-based architectures is expected to bring further gains, especially 
in preparation for the increased complexity and data rates of the HL-LHC Phase--2 upgrade.

\bibliographystyle{unsrtnat}
\bibliography{refs}

@article{CMS:2017wtu,
    author = "Sirunyan, A. M. and others",
    collaboration = "CMS",
    title = "{Identification of heavy-flavour jets with the CMS detector in pp collisions at 13 TeV}",
    eprint = "1712.07158",
    archivePrefix = "arXiv",
    primaryClass = "physics.ins-det",
    reportNumber = "CMS-BTV-16-002, CERN-EP-2017-326",
    doi = "10.1088/1748-0221/13/05/P05011",
    journal = "JINST",
    volume = "13",
    number = "05",
    pages = "P05011",
    year = "2018"
}

@article{Bols_2020,
doi = {10.1088/1748-0221/15/12/P12012},
url = {https://dx.doi.org/10.1088/1748-0221/15/12/P12012},
year = {2020},
month = {dec},
publisher = {},
volume = {15},
number = {12},
pages = {P12012},
author = {E. Bols and J. Kieseler and M. Verzetti and M. Stoye and A. Stakia},
title = {Jet flavour classification using DeepJet},
journal = {Journal of Instrumentation},
}

@article{Qu:2019gqs,
    author = "Qu, Huilin and Gouskos, Loukas",
    title = "{ParticleNet: Jet Tagging via Particle Clouds}",
    eprint = "1902.08570",
    archivePrefix = "arXiv",
    primaryClass = "hep-ph",
    doi = "10.1103/PhysRevD.101.056019",
    journal = "Phys. Rev. D",
    volume = "101",
    number = "5",
    pages = "056019",
    year = "2020"
}

@article{CMS:2024aqx,
    author = "Hayrapetyan, A. and others",
    collaboration = "CMS",
    title = "{Performance of the CMS high-level trigger during LHC Run 2}",
    eprint = "2410.17038",
    archivePrefix = "arXiv",
    primaryClass = "physics.ins-det",
    reportNumber = "CMS-TRG-19-001, CERN-EP-2024-259",
    doi = "10.1088/1748-0221/19/11/P11021",
    journal = "JINST",
    volume = "19",
    number = "11",
    pages = "P11021",
    year = "2024"
}

@techreport{CMS-DP-2017-005,
      collaboration = "CMS",
       title         = "{Heavy flavor identification at CMS with deep neural
                       networks}",
      year          = "2017",
      url           = "https://cds.cern.ch/record/2255736",
      AUTHOR      = "{CMS Collaboration}",
      TYPE        = "CMS Detector Performance Summary",
      NUMBER      = "CMS-DP-2017-005",
      institution = "CERN"
}

@techreport{CMS-DP-2022-030,
      collaboration = "CMS",
      title         = "{Expected Performance of Run-3 HLT b-quark jet
                       identification}",
      year          = "2022",
      url           = "https://cds.cern.ch/record/2825704",
      AUTHOR      = "{CMS Collaboration}",
      TYPE        = "CMS Detector Performance Summary",
      NUMBER      = "CMS-DP-2022-030",
      institution = "CERN"
}

@techreport{CMS-DP-2023-021,
      collaboration = "CMS",
      title         = "{Performance of the ParticleNet tagger on small and
                       large-radius jets at High Level Trigger in~Run~3}",
      year          = "2023",
      url           = "https://cds.cern.ch/record/2857440",
      AUTHOR      = "{CMS Collaboration}",
      TYPE        = "CMS Detector Performance Summary",
      NUMBER      = "CMS-DP-2023-021",
      institution = "CERN"
}

@techreport{CMS-DP-2025-013,
      collaboration = "CMS",
      title         = "{Performance of jet b-tagging in the CMS High-Level
                       Trigger in 2024}",
      year          = "2025",
      url           = "https://cds.cern.ch/record/2931378",
      AUTHOR      = "{CMS Collaboration}",
      TYPE        = "CMS Detector Performance Summary",
      NUMBER      = "CMS-DP-2025-013",
      institution = "CERN"
}

@techreport{CMS-DP-2025-009,
      collaboration = "CMS",
      title         = "{Performance of the ParticleNet b and bb-tagging
                       algorithms in the CMS High-Level Trigger in Run 3}",
      year          = "2025",
      url           = "https://cds.cern.ch/record/2926357",
      AUTHOR      = "{CMS Collaboration}",
      TYPE        = "CMS Detector Performance Summary",
      NUMBER      = "CMS-DP-2025-009",
      institution = "CERN"
}

@article{CMS:2017yfk,
    author = "Sirunyan, A. M. and others",
    collaboration = "CMS",
    title = "{Particle-flow reconstruction and global event description with the CMS detector}",
    eprint = "1706.04965",
    archivePrefix = "arXiv",
    primaryClass = "physics.ins-det",
    reportNumber = "CMS-PRF-14-001, CERN-EP-2017-110",
    doi = "10.1088/1748-0221/12/10/P10003",
    journal = "JINST",
    volume = "12",
    number = "10",
    pages = "P10003",
    year = "2017"
}

@article{Bruschini:2025qba,
    author = "Bruschini, D.",
    collaboration = "CMS",
    title = "{CMS track reconstruction performance and tracking developments during Run 3}",
    doi = "10.1393/ncc/i2025-25080-7",
    journal = "Nuovo Cim. C",
    volume = "48",
    number = "3",
    pages = "80",
    year = "2025"
}

@article{CMS:2018nsn,
    author = "Sirunyan, A. M. and others",
    collaboration = "CMS",
    title = "{Observation of Higgs boson decay to bottom quarks}",
    eprint = "1808.08242",
    archivePrefix = "arXiv",
    primaryClass = "hep-ex",
    reportNumber = "CMS-HIG-18-016, CERN-EP-2018-223",
    doi = "10.1103/PhysRevLett.121.121801",
    journal = "Phys. Rev. Lett.",
    volume = "121",
    number = "12",
    pages = "121801",
    year = "2018"
}

@article{10.1145/3326362,
author = {Wang, Yue and Sun, Yongbin and Liu, Ziwei and Sarma, Sanjay E. and Bronstein, Michael M. and Solomon, Justin M.},
title = {Dynamic Graph CNN for Learning on Point Clouds},
year = {2019},
issue_date = {October 2019},
publisher = {Association for Computing Machinery},
address = {New York, NY, USA},
volume = {38},
number = {5},
issn = {0730-0301},
url = {https://doi.org/10.1145/3326362},
doi = {10.1145/3326362},
journal = {ACM Trans. Graph.},
month = oct,
articleno = {146},
numpages = {12}
}

@article{CMS:2024phk,
    author = "Hayrapetyan, Aram and others",
    collaboration = "CMS",
    title = "{Searches for Higgs boson production through decays of heavy resonances}",
    eprint = "2403.16926",
    archivePrefix = "arXiv",
    primaryClass = "hep-ex",
    reportNumber = "CMS-B2G-23-002, CERN-EP-2024-062",
    doi = "10.1016/j.physrep.2024.09.004",
    journal = "Phys. Rept.",
    volume = "1115",
    pages = "368--447",
    year = "2025"
}

@article{CMS:2020zge,
    author = "Sirunyan, Albert M and others",
    collaboration = "CMS",
    title = "{Inclusive search for highly boosted Higgs bosons decaying to bottom quark-antiquark pairs in proton-proton collisions at $\sqrt{s} =$ 13 TeV}",
    eprint = "2006.13251",
    archivePrefix = "arXiv",
    primaryClass = "hep-ex",
    reportNumber = "CMS-HIG-19-003, CERN-EP-2020-107",
    doi = "10.1007/JHEP12(2020)085",
    journal = "JHEP",
    volume = "12",
    pages = "085",
    year = "2020"
}

@article{CMS:2016fse,
    author = "Khachatryan, Vardan and others",
    collaboration = "CMS",
    title = "{Search for heavy resonances decaying into a vector boson and a Higgs boson in final states with charged leptons, neutrinos, and b quarks}",
    eprint = "1610.08066",
    archivePrefix = "arXiv",
    primaryClass = "hep-ex",
    reportNumber = "CMS-B2G-16-003, CERN-EP-2016-226",
    doi = "10.1016/j.physletb.2017.02.040",
    journal = "Phys. Lett. B",
    volume = "768",
    pages = "137--162",
    year = "2017"
}

@article{PhysRevLett.129.081802,
  author = "Sirunyan, A. M. and others",
  title = {Search for Higgs Boson Pair Production in the Four $b$ Quark Final State in Proton-Proton Collisions at $\sqrt{s}=13\text{ }\text{ }\mathrm{TeV}$},
  collaboration = {CMS Collaboration},
  journal = {Phys. Rev. Lett.},
  volume = {129},
  issue = {8},
  pages = {081802},
  numpages = {20},
  year = {2022},
  month = {Aug},
  publisher = {American Physical Society},
  doi = {10.1103/PhysRevLett.129.081802},
  url = {https://link.aps.org/doi/10.1103/PhysRevLett.129.081802}
}

@article{PhysRevLett.122.121803,
  author = "Sirunyan, A. M. and others",
  title = {Combination of Searches for Higgs Boson Pair Production in Proton-Proton Collisions at $\sqrt{s}=13\text{ }\text{ }\mathrm{TeV}$},
  collaboration = {CMS Collaboration},
  journal = {Phys. Rev. Lett.},
  volume = {122},
  issue = {12},
  pages = {121803},
  numpages = {18},
  year = {2019},
  month = {Mar},
  publisher = {American Physical Society},
  doi = {10.1103/PhysRevLett.122.121803},
  url = {https://link.aps.org/doi/10.1103/PhysRevLett.122.121803}
}

@article{CMS:2018hnq,
    author = "Sirunyan, Albert M and others",
    collaboration = "CMS",
    title = "{Search for $ \mathrm{t}\overline{\mathrm{t}}\mathrm{H} $ production in the $ \mathrm{H}\to \mathrm{b}\overline{\mathrm{b}} $ decay channel with leptonic $ \mathrm{t}\overline{\mathrm{t}} $ decays in proton-proton collisions at $ \sqrt{s}=13 $ TeV}",
    eprint = "1804.03682",
    archivePrefix = "arXiv",
    primaryClass = "hep-ex",
    reportNumber = "CMS-HIG-17-026, CERN-EP-2018-065",
    doi = "10.1007/JHEP03(2019)026",
    journal = "JHEP",
    volume = "03",
    pages = "026",
    year = "2019"
}

@article{CMS:2023tfj,
    author = "Hayrapetyan, Aram and others",
    collaboration = "CMS",
    title = "{Measurement of the Higgs boson production via vector boson fusion and its decay into bottom quarks in proton-proton collisions at $ \sqrt{s} $ = 13 TeV}",
    eprint = "2308.01253",
    archivePrefix = "arXiv",
    primaryClass = "hep-ex",
    reportNumber = "CMS-HIG-22-009, CERN-EP-2023-110",
    doi = "10.1007/JHEP01(2024)173",
    journal = "JHEP",
    volume = "01",
    pages = "173",
    year = "2024"
}

@article{Marchegiani:2024yhi,
    author = "Marchegiani, Matteo",
    collaboration = "CMS",
    title = "{Measurements of the ttH+tH production at CMS}",
    reportNumber = "CMS-CR-2024-298",
    doi = "10.22323/1.478.0133",
    journal = "PoS",
    volume = "LHCP2024",
    pages = "133",
    year = "2025"
}


\end{document}